# Differentially Private Spatiotemporal Trajectory Synthesis with Retained Data Utility

Yuqing Ge
*School of Computer Science and Technology*
*Jiangsu Normal University*
Xuzhou, China
etherious72@163.com

Yunsheng Wang
*School of Computer Science and Technology*
*Jiangsu Normal University*
Xuzhou, China
2386295477@qq.com

Nana Wang(✉)
*School of Computer Science and Technology*
*Jiangsu Normal University*
Xuzhou, China
wangnana_5@aliyun.com

*Abstract*—Spatiotemporal trajectories collected from GPS-enabled devices are of vital importance to many applications, such as urban planning and traffic analysis. Due to the privacy leakage concerns, many privacy-preserving trajectory publishing methods have been proposed. However, most of them could not strike a good balance between privacy protection and good data utility. In this paper, we propose DP-STTS, a differentially private spatiotemporal trajectory synthesizer with high data utility, which employs a model composed of a start spatiotemporal cube distribution and a 1-order Markov process. Specially, DP-STTS firstly discretizes the raw spatiotemporal trajectories into neighboring cubes, such that the model size is limited and the model's tolerance for noise could be enhanced. Then, a Markov process is utilized for the next location point picking. After adding noise under differential privacy (DP) to the model, synthetic trajectories that preserve essential spatial and temporal characteristics of the real trajectories are generated from the noisy model. Experiments on one real-life dataset demonstrate that DP-STTS provides good data utility. Our code is available at https://github.com/Etherious72/DP-STTS.

I. INTRODUCTION

With the popularity of GPS-enabled devices, a growing number of individual trajectories have been generated and collected. Trajectory data can benefit the public in many applications, such as transportation analysis, urban planning, navigation and route recommendation [1-2]. However, due to the sensitive nature of trajectory data, the direct publication of trajectory dataset could potentially expose individual privacy. This emphasizes the need for trajectory privacy protection methods, which protect individual privacy and preserve data utility simultaneously.

Traditional trajectory privacy protection techniques have mostly relied on the idea of *k*-anonymity [3] and its extensions [4], fake location generation [5], sensitive region calculation [6], or new trajectory generation using deep learning methods [7]. Although they can protect the individual privacy to some extent, they are unable to provide tight privacy protection proofs, and may be vulnerable to background knowledge attacks, such as linking attacks and probabilistic attacks [2, 4].

Differential privacy (DP) [8] is a model that can provide provable and strong privacy guarantees. The main idea of differential privacy is to add noise to a dataset so that the query results on the dataset does not change observably when one specific record is included in the dataset or not [9]. Currently, the differentially private trajectory publishing methods are mostly designed for spatial data (i.e., the locations of trajectories) [9-14]. The methods for both the spatial and temporal data (i.e., the timestamps of trajectories) are relatively few.

There are some differentially private spatiotemporal trajectory publishing methods, which build the models using clusters [15-16], prefix trees [17-19] or probability distributions [20]. For example, Hua et al. [15] use the exponential mechanism to merge locations at each time point based on trajectory distances, and then employ a noisy counting approach adapted from the Laplace mechanism for trajectory synthesis. Since each time stamp consumes a certain amount of privacy budget, it might not be appropriate for publishing long trajectories. Al-Hussaeni et al. [19] build a prefix tree to organize the spatiotemporal trajectories, with each level being divided into two sublevels: location and timestamp. Because each level takes some privacy budget, the lengths of the generated trajectories are still limited. Also using the prefix tree, Cai et al. [17] propose a method that only adds noise to the nodes in odd levels and predict the node count of even levels by the Markov transition probability. It may still not be able to generate long trajectories.

Deldar and Abadi [20] present DP-MODR, a method that could generate long trajectories by employing three spatial and statistical characteristics for trajectory release: a starting cell distribution, a trajectory length distribution, and a transition cost matrix. However, the data utility may be not desirable. To illustrate our claim, we compare our proposed method DP-STTS with DP-MODR in Fig. 1. The overall temporal visit frequency distribution in the synthetic trajectories for each technique has been compared to the true dataset (Taxi-1 [21]). The frequencies of visits for each 15 minutes have been calculated. The red curve and the histogram show the overall temporal visit frequency distributions of the true dataset and the synthetic dataset, respectively. Clearly, DP-STTS provides better utility visually. We also demonstrate its superiority quantitatively in Section IV.

This paper presents DP-STTS, a **d**ifferentially **p**rivate **s**patio**t**emporal **t**rajectory **s**ynthesizer with high data utility. We discretize the raw spatiotemporal trajectories into neighboring spatiotemporal cubes, and model them by employing a start cube distribution and a Markov process (Fig. 2). The start cell distribution is used to depict the start position and time for each trajectory, and the Markov process is used for the subsequent location points picking. By adding noise under differential privacy to this model, our method can provide provable privacy guarantees. Our experiments on one real-world dataset show that DP-STTS provides good data utility.

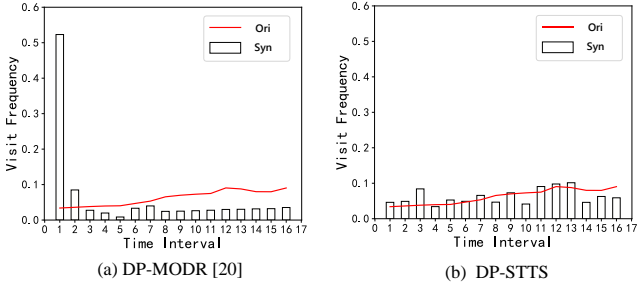

Fig. 1. Overall temporal visit frequency distribution of different methods over Taxi-1($\varepsilon$ =1): (a) DP-MODR [20], and (b) DP-STTS (our model).

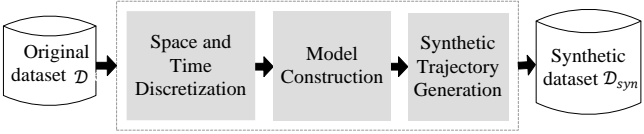

Fig. 2. DP-STTS Framework Overview.

## II. PRELIMINARIES

### A. Notation

Let $\mathcal{D}= \{tr_i|\ i=0,1,…,|\mathcal{D}|-1\}$ be a raw spatiotemporal trajectory dataset with $|\mathcal{D}|$ trajectories, and $tr_i$ denote the *i*-th trajectory of $\mathcal{D}$.

**Definition 1** (Raw trajectory [1]). A raw trajectory $tr^i$ =$\{\ \sigma_j^i(x_j^i,\ y_j^i,\ t_j^i\ )|\ j=0,1,…,|tr^i|-1\ \}$ is a sequence of $|tr^i|$ location points, with each location point $\sigma_j^i(x_j^i, y_j^i, t_j^i)$ corresponding to the location $(x_j^i, y_j^i)$ at timestamp $t_j^i$ ($t_j^i < t_{j+1}^i$).

**Definition 2** (1-order Markov process [10]). A trajectory $tr = \sigma_0\sigma_1 … \sigma_{|tr|-1}$ is said to follow a 1-order Markov process if for every $1 \leq j < |tr|-1$, we have

$$Pr[\sigma_{j+1} = \sigma\ |\ \sigma_0 \cdots \sigma_j] = Pr[\sigma_{j+1} = \sigma\ |\ \sigma_j]. \quad (1)$$

The probability $Pr[\sigma_{j+1} = \sigma\ |\ \sigma_j]$ is a transition probability of the 1-order Markov process. It can be estimated by

$$Pr[\sigma_{j+1} = \sigma\ |] = \frac{\eta(\varpi\sigma,\mathcal{D})}{\eta(\varpi,\mathcal{D})}, \quad (2)$$

where $\varpi$ denotes any symbol, $\eta(\varpi, \mathcal{D})$ is the total number of occurrences of $\varpi$ in $\mathcal{D}$, and $\varpi\sigma$ represents a 2-gram obtained by appending $\sigma$ to $\varpi$.

To make it short, 1-order Markov process is called Markov process for the rest of this paper.

### B. Differential Privacy

Differential privacy [8] is a well-known notion that aims to protect the individuals' privacy when datasets are published. Given two neighboring datasets $\mathcal{D}_1$ and $\mathcal{D}_2$ (i.e., $\mathcal{D}_1$ and $\mathcal{D}_2$ differ in only one record), a DP mechanism bounds the probability difference of obtaining the same answer from $\mathcal{D}_1$ and $\mathcal{D}_2$. That is, the attacker's capability to infer a specific individual's participation in the dataset is restricted, and thus the privacy can be protected.

In our setting, each record of a dataset is a trajectory, and $\mathcal{D}_1$ is obtained by removing/adding a user's record from/into $\mathcal{D}_2$.

**Definition 3** (Differential privacy [8]). A randomized algorithm $M$ is said to be differentially private if for any two neighboring databases $\mathcal{D}_1$ and $\mathcal{D}_2$ ($\mathcal{D}_1, \mathcal{D}_2 \in \mathbb{D}$), and for any possible anonymized output database $\Omega$ ($\Omega \in Range(M)$),

$$Pr[M(\mathcal{D}_1) = \Omega] \leq e^\varepsilon \times Pr[M(\mathcal{D}_2) = \Omega], \quad (3)$$

where $\varepsilon$ is the privacy budget, which defines the privacy level of the algorithm, $\mathbb{D}$ denotes the domain of $M$, and $Range(M)$ represents the set of all possible outputs of $M$.

The Laplace mechanism [8] is one common approach to achieve differential privacy. Suppose $\phi$ is function that maps a dataset $D$ ($D \in \mathbb{D}$) to a vector of $d$ reals, i.e., $\phi: \mathbb{D} \to R^d$. The magnitude of the added Laplace noise depends on the global sensitivity of $\phi$.

**Definition 4** (Global sensitivity [8]). For any function $\phi: D \to R^d$ and any neighboring datasets $\mathcal{D}_1$ and $\mathcal{D}_2$, the sensitivity of $\phi$ is

$$\Delta\phi = \max_{\mathcal{D}_1,\mathcal{D}_2}\|\phi(\mathcal{D}_1) - \phi(\mathcal{D}_2)\|_1. \quad (4)$$

**Theorem 1** (Laplace mechanism [8]). *For any function $\phi: \mathbb{D} \to R^d$, the randomized algorithm M satisfies ε-differential privacy if its output is calculated by adding independently Laplace noise Lap ($\Delta\phi/\varepsilon$) to $\phi(D)$:*

$$M(D) = \phi(D) + Lap\ (\Delta\phi/\varepsilon), \quad (5)$$

*where Lap*($\lambda$) *is a random variable drawn from the Laplace distribution with probability density function* $Pr(x|\lambda) = \frac{1}{2\lambda}e^{\frac{-|x|}{\lambda}}$.

## III. OUR METHOD

The schematic overview of DP-STTS is given in Fig. 2. DP-STTS has three core components: space and time discretization, model construction, and synthetic trajectory generation.

### A. Space and Time Discretization

In our method, we assume the raw spatiotemporal trajectory dataset $\mathcal{D}$ and its neighboring datasets have the same spatial domain $\mathcal{S}$ (e.g., a region) and the same temporal domain $\mathcal{T}$ (e.g., 4:00 -6:00).

Both $\mathcal{S}$ and $\mathcal{T}$ are continuous domains. To limit the size of the model which we will build in Section III.B, we discretize the trajectories of $\mathcal{D}$ into neighboring cubes (Fig. 3). We first impose a $g_h \times g_w$ uniform grid over $\mathcal{S}$ and evenly divide $\mathcal{T}$ into $g_t$ time intervals to get a cube set $\mathcal{ST}$ =$\{\ st_i(st_i.x, st_i.y, st_i.t)|i = 0,1,…,|\mathcal{ST}|-1\ \}$ ($|\mathcal{ST}| = g_h \times g_w \times g_t$), where $st_i$ denotes the *i*-th cube of $\mathcal{ST}$, and $st_i.x$, $st_i.y$, $st_i.t$ represent the longitude index, the latitude index and time index within the spatiotemporal domain defined by $\mathcal{S}$ and $\mathcal{T}$, respectively. Then, a cube trajectory dataset $\mathcal{D}_c$ is obtained by transforming each location point $\sigma_j^i(x_j^i, y_j^i, t_j^i)$ into its cube form $\sigma_{c,j}^i(x_{c,j}^i, y_{c,j}^i, t_{c,j}^i)$ ($\sigma_{c,j}^i \in \mathcal{ST}$),

$$\begin{cases} x_{c,j}^i = \lfloor (x_j^i - left)/d_w \rfloor \\ y_{c,j}^i = \lfloor (y_j^i - bottom)/d_h \rfloor, \\ t_{c,j}^i = \lfloor (t_j^i - sTime)/d_t \rfloor \end{cases} \quad (6)$$

where

$$\begin{cases} d_w = (right - left)/g_w \\ d_h = (top - bottom)/g_h \\ d_t = (eTime - sTime)/g_t \end{cases} \quad (7)$$

and $P_1$(*left*, *bottom*) and $P_2$ (*right*, *top*) are the lower left point and the upper right point that define the minimum encasing rectangle of $\mathcal{S}$, and $sTime$ and $eTime$ are the earliest time and the latest time of $\mathcal{T}$, respectively. After this step, $tr^i = \{\sigma_j^i(x_j^i, y_j^i, t_j^i)|\ j = 0,1, \ldots, |tr^i| - 1\}$ becomes a cube trajectory $tr_c^i = \{\sigma_{c,j}^i (x_{c,j}^i, y_{c,j}^i, t_{c,j}^i)|\ j = 0,1, \ldots, |tr_c^i| - 1\}$. We add a stopping symbol '^' to each cube trajectory. For simplicity, we use one cube to represent the adjacent same cubes of a cube trajectory.

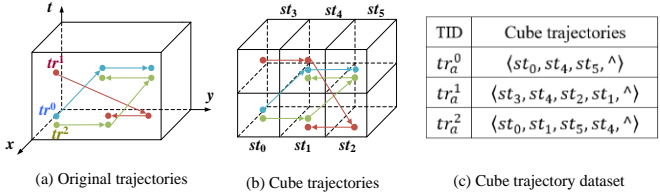

Fig. 2. (a) Original trajectories, (b) cube trajectories corresponding to Fig.3 (a), and (c) cube trajectory dataset corresponding to Fig.3 (b).

Then, we can use the transitions between cubes to simulate the moves between the location points of $\mathcal{D}$. For each cube, we only consider the transitions from itself to its *neighboring cubes* and the stopping symbol '^'. We call a cube $st_i(st_i.x, st_i.y, st_i.t)$ a *neighboring cube* of $st_j(st_j.x, st_i.y, st_j.t)$, if $st_i$ satisfies one of following two sets of conditions:

$$\begin{cases} |st_i.x - st_j.x| \leq 1 \\ |st_i.y - st_j.y| \leq 1 \\ 0 \leq st_i.t - st_j.t \leq 1 \\ st_i \neq st_j \end{cases} \text{or} \begin{cases} st_i.t - st_j.t = v, \\ st_i = st_j \end{cases},$$

where $v$ ( $v = 1, 2, \ldots$) is parameter for time discretization. Let $neighbor(st_i)$ be the *neighboring cube* set of $st_i$. For every two adjacent cubes $\sigma_{c,j}^i$ and $\sigma_{c,j+1}^i$ of $tr_c^i$, we insert cubes between them based on interpolation, if $\sigma_{c,j+1}^i$ is not a *neighboring cube* of $\sigma_{c,j}^i$. Then, $tr_c^i$ becomes $tr_a^i = \{\sigma_{a,j}^i (x_{a,j}^i, y_{a,j}^i, t_{a,j}^i)|\ j = 0,1, \ldots, |tr_a^i| - 1\}$ after cube insertion, where $\sigma_{a,j}^i$ represents the *j*-th cube of $tr_a^i$. Let $\mathcal{D}_a = \{tr_a^i|\ i = 0,1,\ldots, |\mathcal{D}_a|-1\}$ ($|\mathcal{D}_a| = |\mathcal{D}|$) be the obtained cube trajectory dataset.

### B. Model Construction

In this phase, we organize the trajectories of $\mathcal{D}_a$ by combining a start cube distribution and a Markov process, and add noise to the model in a differentially private way. Let $\varepsilon_s$ and $\varepsilon_m$ be the privacy budgets allocated to start cube distribution and the Markov process, respectively. The whole privacy budget $\varepsilon$ is shared by $\varepsilon_s$ and $\varepsilon_m$, based on a parameter $\delta$. We set $\varepsilon_s = \delta\varepsilon$, $\varepsilon_m = (1 - \delta)\varepsilon, 0 < \delta < 1$.

*B.1 Noisy Start Cube Distribution*

Let $n(st_i)(i = 0, 1, \ldots, |\mathcal{ST}| - 1 )$ be the number of trajectories which start from cube $st_i$. We regard the set $N = \{n(st_i)|\ i = 0, 1, \ldots, |\mathcal{ST}| - 1 \}$ as the start cube distribution. Let $N' = \{n(st_i)'|\ i = 0, 1, \ldots, |\mathcal{ST}| - 1 \}$ be the noisy start cube distribution. We get $N'$ using the following two steps:

**Step 1:** Add Laplace noise to $n(st_i)$ by

$$n(st_i)' = n(st_i) + Lap(\Delta n(\cdot)/\varepsilon_s), \quad (8)$$

where $\Delta n(\cdot)$ is the sensitivity of function $n(\cdot)$. If a trajectory is added to or removed from $\mathcal{D}_a$, the value of $|n(\cdot)|$ changes at most one. Thus, we have $\Delta n(\cdot) = 1$.

**Step 2:** Enforce consistency constraints among all the elements of $N'$ by

$$n(st_i)' = \frac{n(st_i)'}{\sum_{j=0}^{|\mathcal{ST}|-1} n(st_j)'}. \quad (9)$$

**Theorem 2.** *Noisy start cube distribution calculation satisfies $\varepsilon_s$-DP.*

*Proof.* The noisy start cube calculation consists of two steps: noise addition and consistency constraints enforcement. Because the consistency constraint enforcement is a postprocessing step [22], we just need to prove $\frac{Pr(M_s(\mathcal{D})=N\prime)}{Pr(M_s(\widetilde{\mathcal{D}})=N\prime)} \leq e^{\varepsilon_s}$, where $M_s$ is the noise addition step, and $\widetilde{\mathcal{D}}$ denotes a neighboring dataset of $\mathcal{D}$. Let's assume $\mathcal{D}$ and $\widetilde{\mathcal{D}}$ become $\mathcal{D}_a$ and $\widetilde{\mathcal{D}}_a$ after the space and time discretization step. $\mathcal{D}_a$ and $\widetilde{\mathcal{D}}_a$ are also two neighboring datasets. Therefore, we have $\frac{Pr(M_s(\mathcal{D})=N\prime)}{Pr(M_s(\widetilde{\mathcal{D}})=N\prime)} = \frac{Pr(M_s(\mathcal{D}_a)=N\prime)}{Pr(M_s(\widetilde{\mathcal{D}}_a)=N\prime)}$.

Let's assume a trajectory $tr^e$ of $\mathcal{D}$ starts from $st_j$. Then, $tr^e$ only affects $n(st_j)$. Therefore, we have

$$\frac{Pr(M_s(\mathcal{D}_a) = N')}{Pr(M_s(\widetilde{\mathcal{D}}_a) = N')} = \prod_{st_i \in \mathcal{ST}} \frac{\exp\left(-\varepsilon_s \frac{|n(st_i)' - n(st_i)|}{\Delta n(\cdot)}\right)}{\exp\left(-\varepsilon_s \frac{|n(st_i)' - n(\widetilde{st}_i)|}{\Delta n(\cdot)}\right)}$$

$$\leq \exp(\varepsilon_s|n(st_j) - n(\widetilde{st}_j)|)$$

$$\leq e^{\varepsilon_s}. \quad (10)$$

That is, the noisy start cube calculation satisfies $\varepsilon_s$-DP.

*B.2 Noisy Markov Process*

We represent a Markov process by using a transition matrix $TM$ (as shown in Fig. 4). We first calculate a frequency matrix $FM$ by scanning all the trajectories of $\mathcal{D}_a$. After adding noise to $FM$, $TM$ is derived from the noisy $FM$ (denoted by $FM'$).

$FM$ is a matrix that has $|\mathcal{ST}|$ rows and $(|\mathcal{ST}| + 1)$ columns. The *i*-th row of $FM$ is associated with a cube $st_i$ ($i = 0, 1, \ldots, |\mathcal{ST}| - 1$), and the *j*-th column of $FM$ is associated with a symbol $c_j$, where $c_j$ is a cube $st_j \in \{\mathcal{ST}\}$ or the stopping symbol '^'. Let $FM_{i,j}$ be the element on the $i$-th row and the $j$-th column of $FM$, which stores the frequency of a 2-gram $st_i c_j$.

|  | $st_0$ | $st_1$ | $st_2$ | $st_3$ | $st_4$ | $st_5$ | ^ |
|---|---|---|---|---|---|---|---|
| $st_0$ | 0 | 1/4 | 0 | 1/3 | 0 | 0 | 0 |
| $st_1$ | 0 | 0 | 0 | 0 | 1/4 | 1/4 |  |
| $st_2$ | 0 | 1/4 | 0 | 0 | 0 | 0 | 0 |
| $st_3$ | 0 | 0 | 0 | 1/4 | 0 | 0 | 0 |
| $st_4$ | 0 | 1/4 | 0 | 0 | 1/3 | 1/4 |  |
| $st_5$ | 0 | 0 | 0 | 0 | 1/4 | 0 | 1/3 |

(a) $FM$

|  | $st_0$ | $st_1$ | $st_2$ | $st_3$ | $st_4$ | $st_5$ | ^ |
|---|---|---|---|---|---|---|---|
| $st_0$ | 0 | 1/2 | 0 | 1/2 | 0 | 0 | 0 |
| $st_1$ | 1/2 | 0 | 0 | 0 | 1/2 | 0 |  |
| $st_2$ | 0 | 1/2 | 0 | 0 | 0 | 0 | 0 |
| $st_3$ | 0 | 0 | 0 | 0 | 0 | 0 | 0 |
| $st_4$ | 0 | 0 | 1/2 | 1/2 | 0 | 0 | 1 |
| $st_5$ | 0 | 0 | 0 | 0 | 1/2 | 0 | 1/2 |

(b) $FM'$

|  | $st_0$ | $st_1$ | $st_2$ | $st_3$ | $st_4$ | $st_5$ | ^ |
|---|---|---|---|---|---|---|---|
| $st_0$ | 0 | 1/2 | 0 | 1/2 | 0 | 0 | 0 |
| $st_1$ | 1/2 | 0 | 0 | 0 | 1/2 | 0 |  |
| $st_2$ | 0 | 1 | 0 | 0 | 0 | 0 | 0 |
| $st_3$ | 0 | 0 | 0 | 0 | 0 | 0 | 0 |
| $st_4$ | 0 | 0 | 1/4 | 1/4 | 0 | 0 | 1/2 |
| $st_5$ | 0 | 0 | 0 | 0 | 1/2 | 0 | 1/2 |

(c) $TM$

Fig. 4. (a) The frequency matrix $FM$ of the cube trajectory dataset in Fig. 3(c), (b) an example of the noisy frequency matrix $FM'$ of Fig. 4(a), and (c) the transition matrix $TM$ derived from Fig. 4(b).

Since the elements of a trajectory are neighboring cubes, we only need to consider the transitions between neighboring cubes. We compute $FM$, add noise to it and then derive $TM$ as follows.

**Step1:** Create two matrices $FM = \{FM_{i,j} = 0 \mid i = 0,1,...,|\mathcal{ST}|-1, j = 0,1,...,|\mathcal{ST}|\}$ and $FM' = \{FM_{i,j}' = 0 \mid i = 0,1,...,|\mathcal{ST}|-1, j = 0,1,...,|\mathcal{ST}|\}$.

**Step2:** Calculate the frequency of every 2-gram of $\mathcal{D}_a$. For a trajectory $tr_a^i = \{\sigma_{a,j}^i \mid j = 0,1,...,|tr_a^i|-1\}$, we scan its cubes from $\sigma_{a,0}^i$. For the 2-gram $st_i n_j$, we compute its frequency $FM_{i,j}$ by a query $\psi(\cdot)$ over $\mathcal{D}_a$,

$$FM_{i,j} = \psi(st_i c_j, \mathcal{D}_a) = \sum_{tr_a^i \in \mathcal{D}_a} \frac{f(st_i c_j, tr_a^i)}{|tr_a^i|-1}, \quad (11)$$

where $f(st_i c_j, tr_a^i)$ denotes the occurrence of $st_i c_j$ in $tr_a^i$.

**Step3:** Add Laplace noise to each element $FM_{i,k}$ ($k$ satisfies that $c_k \in \{neighbor(st_i) \cup \{^\wedge\}\}$) of $FM$ to get $FM'$ by $\{\mathcal{ST} \cup \{^\wedge\}\}$

$$FM_{i,k}' = FM_{i,k} + Lap(\Delta\psi(\cdot)/\varepsilon_m), \quad (12)$$

where $\Delta\psi(\cdot)$ is the sensitivity of $\psi(\cdot)$. The value of $|\psi(\cdot)|$ changes at most one if a trajectory is removed from or added to $\mathcal{D}_a$. Thus, we have $\Delta\psi(\cdot) = 1$.

**Step4:** Create the transition matrix $TM = \{TM_{i,j} = 0 \mid i = 0,1,...,|FM|-1, j = 0,1,...,|\mathcal{ST}|\}$, and calculate $TM_{i,k}$ ($k$ satisfies that $c_k \in \{neighbor(st_i) \cup \{^\wedge\}\}$) by

$$TM_{i,k} = FM_{i,k}' / \sum_{u=0}^{|\mathcal{ST}|} FM_{i,u}'. \quad (13)$$

An example of a Markov process for the cube trajectory dataset in Fig. 3(c) is shown in Fig. 4. After adding noise, an example of the noisy frequency matrix $FM'$ of Fig. 4(a) is shown in Fig. 4(b). Fig. 4(c) gives the transition matrix $TM$ derived from Fig. 4(b).

**Theorem 3.** *Noisy Markov process calculation satisfies $\varepsilon_m$-DP.*

*Proof.* Noisy frequency matrix calculation (Step 1- Step 3) and transition matrix calculation (Step 4) are the two parts of the noisy Markov process calculation. The calculation of transition matrix is a postprocessing step, which consumes zero privacy budget. Thus, similar to the proof of Theorem 2, we only need to prove $\frac{Pr(M_p(\mathcal{D}_a)=FM')}{Pr(M_p(\widetilde{\mathcal{D}}_a)=FM')} \leq e^{\varepsilon_m}$, where $M_p$ denotes the noisy frequency matrix calculation.

A trajectory $tr_a^i$ of $\mathcal{D}_a$ has at most $(|tr_a^i|-1)$ different kinds of 2-grams, which means it affects at most $(|tr_a^i|-1)$ elements of $FM$. Let $\alpha$ denote the set of elements of $FM$ that are affected by $tr_a^i$, and $\widetilde{FM}$ be the frequency matrix of $\widetilde{\mathcal{D}}_a$. We have

$$\frac{Pr(M_p(\mathcal{D}_a)=FM')}{Pr(M_p(\widetilde{\mathcal{D}}_a)=FM')} = \prod_{FM_{i,j} \in FM} \frac{\exp\left(-\varepsilon_m \frac{|FM_{i,j}'-FM_{i,j}|}{\Delta\psi(\cdot)}\right)}{\exp\left(-\varepsilon_m \frac{|FM_{i,j}'-\widetilde{FM}_{i,j}|}{\Delta\psi(\cdot)}\right)}$$

$$\leq \prod_{FM_{i,j} \in \alpha} \exp(\varepsilon_m |FM_{i,j} - \widetilde{FM}_{i,j}|)$$

$$= \exp(\varepsilon_m \sum_{FM_{i,j} \in \alpha} |FM_{i,j} - \widetilde{FM}_{i,j}|). \quad (14)$$

According to Eq. (11), we can get $\sum_{FM_{i,j} \in \alpha} |FM_{i,j} - \widetilde{FM}_{i,j}| = 1$. Thus, Eq. (14) can be rewritten as

$$\frac{Pr(M_p(\mathcal{D}_a)=FM')}{Pr(M_p(\widetilde{\mathcal{D}}_a)=FM')} \leq \exp(\varepsilon_m \sum_{FM_{i,j} \in \alpha} |FM_{i,j} - \widetilde{FM}_{i,j}|) = e^{\varepsilon_m}. \quad (15)$$

That is, the noisy Markov process calculation satisfies $e^{\varepsilon_m}$-DP.

### C. Synthetic Trajectory Generation

In this step, we generate the synthetic trajectory dataset $\mathcal{D}_{syn}$. For each cube $st_i \in \mathcal{ST}$, $n(st_i)'$ copies of $st_i$ are firstly sampled and appended to the output as its corresponding trajectories. Let $tr' = st_i$ be such a trajectory. According to the last cube $st_i$ of $tr'$, a symbol $c_j$ ($c_j \in \{\mathcal{ST} \cup \{^\wedge\}\}$) is selected as the next cube of $tr'$ with a probability $TM_{i,j}$ and appended to $tr'$. This cube selection and appending process is repeated until the stopping symbol ^ is appended to $tr'$ or the maximum length $len$ is reached.

Let $\sigma_j'(x_j', y_j', t_j')$ be the $j$-th cube of $tr'$. A synthetic trajectory of $\mathcal{D}_{syn}$ is generated by transforming each cube of $tr'$ into a location point $\sigma_{syn,j}(x_{syn,j}, y_{syn,j}, t_{syn,j})$,

$$\begin{cases} x_{syn,j} = left + x_j' \cdot d_w + d_w/2 \\ y_{syn,j} = bottom + y_j' \cdot d_h + d_h/2 \\ t_{syn,j} = sTime + t_j' \cdot d_t + d_t/2 \end{cases}. \quad (16)$$

**Theorem 4.** *DP-STTS satisfies $\varepsilon$-DP.*

*Proof.* Space and time discretization, model construction, and synthetic trajectory generation are the three components of DP-STTS. Zero privacy budget is consumed in the space and time discretization and the synthetic trajectory generation steps. The model construction has two parts: the noisy starting cube calculation and the noisy Markov process calculation, whose whole privacy budget is $\varepsilon = \varepsilon_s + \varepsilon_m$ according to the sequential composition property [23]. Therefore, DP-STTS satisfies $\varepsilon$-DP.

## IV. EXPERIMENTAL RESULTS AND ANALYSIS

### A. Experiment Setup

*A.1 Datasets*

We use the dataset of the Taxi Service Prediction Challenge at ECML-PKDD2015 [21] to evaluate the performance of DP-STTS. This dataset is composed of a complete year (from 01/07/2013 to 30/06/2014) data of 1.72 million traces for 442 taxis in Porto. We extract two sub-datasets from it (namely Taxi-

1 and Taxi-2) using two bounding boxes: ((41.104N, 8.665W), (41.250N, 8.528W)) and ((41.064N, 8.662W), (41.210N, 8.525W)), each of which is 11.49km×16.23km in area. The domain of $\mathcal{T}$ is 14:00-18:00.

*A.2 Utility Metrics*

We evaluate DP-STTS using six quantitative metrics.

- **Overall temporal visit frequency distribution [7].** This metric is used to measure the temporal similarity of different methods. In our experiments, we count the frequencies of visits for each time interval (e.g. 15 minutes) in $\mathcal{D}$ and $\mathcal{D}_{syn}$ and convert them into probability distribution for the overall temporal visit frequency distribution evaluation.

- **Location visit average relative error (Location $AvRE$) [20].** This metric is employed to assess the quality of similarities between locations' popularity. We impose a uniform grid $G$ over the spatial domain $\mathcal{S}$ of $\mathcal{D}$, and regard each cell as a discrete location. Let $\mathcal{L} = \{\mathcal{L}_i | i = 0, 1, ..., \gamma - 1\}$ be the cell set of $G$, and $\gamma$ be the number of $\mathcal{L}$'s elements. The location visit *relative error* (*RE*) of $\mathcal{L}_i$ is

$$RE = \frac{|pop(\mathcal{L}_i, \mathcal{D}) - pop(\mathcal{L}_i, \mathcal{D}_{syn})|}{max\{pop(\mathcal{L}_i, \mathcal{D}), \lambda\}},$$

where $pop(\mathcal{L}_i, \mathcal{D})$ denotes the number of times $\mathcal{L}_i$ is visited by the trajectories of $\mathcal{D}$, $\lambda$ is a *sanity bound* that mitigates the effect of locations with extremely small visits. The location $AvRE$ is calculated by averaging all the *RE* of all the locations of $\mathcal{L}$. In the experiments, we set $\lambda = 0.1\% \times |\mathcal{D}|$, the scale of $G$ is 20×20.

- **Location Kendall-tau coefficient (Location $KT$) [20].** We use this metric to evaluate the similarities and discrepancies between locations' popularity ranking in $\mathcal{D}$ and $\mathcal{D}_{syn}$. We regard the location pair $(\mathcal{L}_i, \mathcal{L}_j)$ as concordant if the popularity ranks of $\mathcal{L}_i$ and $\mathcal{L}_j$ in sorted order agree in $\mathcal{D}$ and $\mathcal{D}_{syn}$. That is, $(\mathcal{L}_i, \mathcal{L}_j)$ is concordant if one of the following conditions holds:

$$(pop(\mathcal{L}_i, \mathcal{D}) > pop(\mathcal{L}_j, \mathcal{D})) \wedge (pop(\mathcal{L}_i, \mathcal{D}_{syn}) > pop(\mathcal{L}_j, \mathcal{D}_{syn}))$$
$$(pop(\mathcal{L}_i, \mathcal{D}) < pop(\mathcal{L}_j, \mathcal{D})) \wedge (pop(\mathcal{L}, \mathcal{D}_{syn}) < pop(\mathcal{L}_j, \mathcal{D}_{syn}))$$

The location $KT$ coefficient is computed by

$$Location\ KT = \frac{(\#of\ concordant\ location\ pairs) - (\#of\ discordant\ location\ pairs)}{\gamma(\gamma-1)/2}.$$

- **The frequent pattern Kendall-tau coefficient (FP KT) [11, 20].** This metric is used to evaluate the similarities and discrepancies between frequent patterns' popularity ranking in $\mathcal{D}$ and $\mathcal{D}_{syn}$. Assume a pattern $P$ is represented by an ordered list of the locations of $\mathcal{L}$. Let $o(P, \mathcal{D})$ be the number of occurrences of $P$ in $\mathcal{D}$, and $FP_G^k(\mathcal{D}) = \{P_i | i = 0, 1, ..., k-1\}$ denote the top-$k$ patterns in $\mathcal{D}$. For two frequent patterns $P_i, P_j \in FP_G^k(\mathcal{D})$ $(i, j = 0, 1, ..., k-1)$ of $FP_U^k(D)$, the pair $(P_i, P_j)$ is concordant if it satisfies one of the following two requirements:

$$(o(P_i, \mathcal{D}) > o(P_j, \mathcal{D})) \wedge (o(P_i, \mathcal{D}_{syn}) > o(P_j, \mathcal{D}_{syn}))$$
$$(o(P_i, \mathcal{D}) < o(P_j, \mathcal{D})) \wedge (o(P_i, \mathcal{D}_{syn}) < o(P_j, \mathcal{D}_{syn}))$$

Then, the frequent pattern $KT$ coefficient can be computed by

$$FP\ KT = \frac{(\#of\ concordant\ FP\ pairs) - (\#of\ discordant\ FP\ pairs)}{k(k-1)/2}.$$

In our experiments, we set $k = 200$, and the length of each pattern is from 2 to 8.

- **Trip error [10-11, 20].** This metric aims to evaluate the preservation ability of the correlations between the start and end locations of the original trajectory dataset. Let $\tau(G, \mathcal{D})$ represent the distribution of all possible (start location, end location) pairs of $\mathcal{D}$. The trip error is computed by: $JSD(\tau(G, \mathcal{D}), \tau(G, \mathcal{D}_{syn}))$, where $JSD$ denotes the Jensen-Shannon divergence.

- **Length error [20].** To assess the preservation ability of the trajectory length distribution, we quantize the trajectory lengths of $\mathcal{D}$ into 20 equal width buckets, and calculate the length error by: $JSD(len(\mathcal{D}), len(\mathcal{D}_{syn}))$, where $len(\mathcal{D})$ is the empirical distribution of the trajectory lengths on $\mathcal{D}$.

*B. Comparison with Related Works*

*B.1 Related Works*

We compare DP-STTS with DP-MODR [20]. In the comparison, we set $g_h \times g_w = 20 \times 20$ and $g_t = 16$ (i.e., the duration of each time interval is 15 minutes) for all both of them. We perform DP-MODR by taking $h_{max} = 3$, and $\varepsilon_1 = \varepsilon_2 = \varepsilon_3 = \varepsilon/3$. DP-STTS is run by taking $\delta = 0.5$, $v = 2$, and $len = 125$. The parameters recommended by the authors are used when applicable. Because the Laplace mechanisms and the exponential mechanisms used in the two methods are probabilistic, we repeat each experiment five times and report the average results.

*B.2 Utility Comparisons*

We compare DP-STTS with DP-MODR [20] with varying privacy budget $\varepsilon$. The results are listed in Table I, where the best result in each category is shown in bold. The overall temporal visit frequency distribution of different methods over Taxi-1 and Taxi-2 are illustrated in Fig. 1 and Fig. 5, respectively.

TABLE I. COMPARING THE PROPOSED METHOD WITH ITS COMPETITOR. BEST RESULT IN EACH CATEGORY IS SHOWN IN BOLD. FOR ALL THE METRICS EXCEPT LOCATION KT, LOWER VALUES ARE BETTER.

| Metric | $\varepsilon$ | Taxi-1 | | Taxi-2 | |
|---|---|---|---|---|---|
| | | [20] | DP-STTS | [20] | DP-STTS |
| Location Visit $AvRE$ | 0.5 | 0.781 | **0.504** | 1.055 | **0.480** |
| | 1 | 0.779 | **0.537** | 1.020 | **0.476** |
| Location $KT$ | 0.5 | 0.533 | **0.683** | 0.639 | **0.704** |
| | 1 | 0.543 | **0.704** | 0.636 | **0.723** |
| FP $KT$ | 0.5 | 0.081 | **0.457** | 0.031 | **0.522** |
| | 1 | 0.069 | **0.463** | 0.030 | **0.498** |
| Trip Error | 0.5 | 0.570 | **0.273** | 0.551 | **0.260** |
| | 1 | 0.570 | **0.272** | 0.552 | **0.258** |
| Length Error | 0.5 | **0.021** | 0.093 | **0.058** | 0.084 |
| | 1 | **0.021** | 0.094 | **0.058** | 0.088 |

We observe that when subjected to the same level of privacy, DP-STTS generally provides higher data utility. Although its length error values are not smaller than that of DP-MODR (the biggest difference is smaller than 0.08), the advancements are significant in terms of Location visit $AvRE$, Location $KT$, FP $KT$, and Trip Error. DP-MODR [20] builds pruned noisy cost-sensitive path trees to simulate the most frequent patterns. Since some less frequent patterns have been neglected, it may be harder to preserve good aggregate properties. In DP-STTS, the neighboring cubes are used represent the trajectories which may limit the model size and enhance the model's resistance against noise. Furthermore, using the Markov process to simulate the transitions between the neighboring cubes may help to preserve more moving patterns of the trajectories.

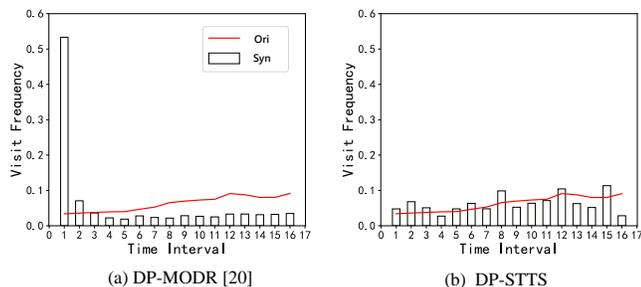

Fig. 5. Overall temporal visit frequency distribution of different methods over Taxi-2 ($\varepsilon=1$): (a) DP-MODR [20], and (b) DP-STTS.

## V. CONCLUSIONS

We present DP-STTS, a differentially private spatiotemporal trajectory synthesizer. It could provide both provable privacy guarantees and good data utility by discretizing the original trajectories into neighboring spatiotemporal cubes, and adding noise under differential privacy to a model which is composed of a start cube distribution and a 1-order Markov process. Experimental results on a real-world dataset show that DP-STTS performs well in terms of the preservation of the essential spatial and temporal characteristics of the real trajectories.